\documentclass[
 reprint,
amsmath,amssymb,aps, floatfix, superscriptaddress
]{revtex4-2}
\usepackage[subpreambles=true]{standalone}
\usepackage{import, inputenc, amsmath, amssymb, amsfonts, caption, graphicx, epstopdf, amsthm, hyperref, subcaption}
\numberwithin{equation}{section}	
\usepackage{fancyhdr,cancel,verbatim,listings, diagbox,float, braket,tensor, tikz,multirow, bm, placeins, pgfplots, pgfplotstable, array,upgreek}
\usepackage[normalem]{ulem} 
\usetikzlibrary{decorations.pathmorphing}
\usepackage[space]{grffile} 		
\usepackage[titletoc]{appendix}
\usetikzlibrary{intersections, arrows,decorations.pathmorphing,snakes, shapes, positioning, patterns, decorations.text,decorations.pathreplacing}

\usetikzlibrary{arrows}
\tikzset{level/.style={thick},
         connect/.style = {thin},
         excite/.style={very thick, blue,->,shorten >=1.5pt,shorten <=1.5pt,>=stealth},
         emit/.style={very thick, red,->,shorten >=1.5pt,shorten <=1.5pt,>=stealth},
         coupling/.style = {line width=2pt, red,->,shorten >=2pt,shorten <=2pt,>=stealth}
}
\usetikzlibrary{shapes.gates.logic.US, calc}
\usepackage[siunitx]{circuitikz}\ctikzset{resistor = european}

\newcommand{\specialcell}[2][c]{%
\begin{tabular}[#1]{@{}c@{}}#2\end{tabular}}

\begin{document}
\title{Rydberg-atom-based single-photon detection for haloscope axion searches}
\begin{abstract}
We propose a Rydberg-atom-based single-photon detector for signal readout in dark matter haloscope experiments between 40 $\upmu$eV and 200 $\upmu$eV (10 GHz and 50 GHz). At these frequencies, standard haloscope readout using linear amplifiers is limited by quantum measurement noise, which can be avoided by using a single-photon detector. Our single-photon detection scheme can offer scan rate enhancements up to a factor of 10$^4$ over traditional linear amplifier readout, and is compatible with many different haloscope cavities. We identify multiple haloscope designs that could use our Rydberg-atom-based single-photon detector to search for QCD axions with masses above 40 $\upmu$eV (10 GHz), currently a minimally explored parameter space.
 
\end{abstract}
\date{\today}

\author{Eleanor Graham}
\affiliation{Department of Physics, Yale University, New Haven, Connecticut 06520, USA}
\affiliation{Wright Laboratory, Department of Physics, Yale University, New Haven, Connecticut 06520, USA}
\author{Sumita Ghosh}
\thanks{now at MIT}
\affiliation{Department of Applied Physics, Yale University, New Haven, Connecticut 06520, USA}
\affiliation{Wright Laboratory, Department of Physics, Yale University, New Haven, Connecticut 06520, USA}
\author{Yuqi Zhu}
\thanks{now at Stanford}
\affiliation{Department of Physics, Yale University, New Haven, Connecticut 06520, USA}
\affiliation{Wright Laboratory, Department of Physics, Yale University, New Haven, Connecticut 06520, USA}
\author{Xiran Bai}
\affiliation{Department of Physics, Yale University, New Haven, Connecticut 06520, USA}
\affiliation{Wright Laboratory, Department of Physics, Yale University, New Haven, Connecticut 06520, USA}
\author{Sidney B. Cahn}
\affiliation{Department of Physics, Yale University, New Haven, Connecticut 06520, USA}
\affiliation{Wright Laboratory, Department of Physics, Yale University, New Haven, Connecticut 06520, USA}
\author{Elsa Durcan}
\affiliation{Department of Physics, Yale University, New Haven, Connecticut 06520, USA}
\affiliation{Wright Laboratory, Department of Physics, Yale University, New Haven, Connecticut 06520, USA}
\author{Michael J. Jewell}
\affiliation{Department of Physics, Yale University, New Haven, Connecticut 06520, USA}
\affiliation{Wright Laboratory, Department of Physics, Yale University, New Haven, Connecticut 06520, USA}
\author{Danielle H. Speller}
\affiliation{Department of Physics and Astronomy, The Johns Hopkins University, 3400 North Charles Street Baltimore, MD, 21211}
\author{Sabrina M. Zacarias}
\affiliation{Department of Physics, Yale University, New Haven, Connecticut 06520, USA}
\affiliation{Wright Laboratory, Department of Physics, Yale University, New Haven, Connecticut 06520, USA}
\author{Laura T. Zhou}
\thanks{now at Stanford}
\affiliation{Department of Physics, Yale University, New Haven, Connecticut 06520, USA}
\affiliation{Wright Laboratory, Department of Physics, Yale University, New Haven, Connecticut 06520, USA}
\author{Reina H. Maruyama}
\affiliation{Department of Physics, Yale University, New Haven, Connecticut 06520, USA}
\affiliation{Wright Laboratory, Department of Physics, Yale University, New Haven, Connecticut 06520, USA}

\maketitle

\section{Introduction}
The axion is a compelling solution to the strong charge–parity ($CP$) problem in quantum chromodynamics (QCD) and a well-motivated dark matter candidate in astrophysics and cosmology~\cite{Peccei, Peccei2, Weinberg1978, Wilczek1978, Preskill1983, Abbott1982, Dine1982}. Though the axion mass $m_a$ is linked to the energy scale of the spontaneous symmetry breaking $f_a$, QCD itself does not constrain $m_a$ or $f_a$~\cite{diCortona2016}. Observations from astrophysics and cosmology constrain  $m_a \sim 10^{-6}-10^3\: \upmu$eV~\cite{Raffelt, Arvanitaki2015, marsh2016, Cardoso2018, Buschmann2022}. The coupling strength between the axion and the standard model depends on the axion mass: for a given value of $m_a$, there are a range of axion-photon couplings $g_{\gamma}$ which are compatible with QCD.  This region is typically spanned by two benchmark QCD axion models, the Kim-Shifman-Vainshtein-Zakharov (KSVZ) model~\cite{Kim1979, SHIFMAN1980} and the Dine-Fischler-Srednicki-Zhitnitsky (DFSZ) model~\cite{DINE1981, Zhitnitsky1980}. Axion-like particles (ALPs) with photon couplings outside the range predicted from QCD can also act as dark matter, although they cannot solve the strong CP problem~\cite{Preskill1983}.

Among laboratory axion searches to date, only microwave cavity haloscope experiments are sensitive to QCD axions \cite{admx2021,Jewell2023,CAPP2021,TASEH2021,QUAX2022}. Haloscope experiments make use of the axion-photon conversion in a strong magnetic field, using a resonant cavity with a high quality factor to collect the signal~\cite{Sikivie1983, Sikivie1985}. The resulting signal is typically amplified with linear amplifiers such as HEMTs or JPAs~\cite{admx2021,Jewell2023}. Even with amplification, the resulting signal rates are very low, necessitating extremely low noise levels in haloscope experiments approaching or surpassing the standard quantum limit (SQL). For instance, in the case of axions with 10 $\upmu$eV mass and KSVZ coupling strength, the expected signal power for an experiment such as HAYSTAC would be on the order of $10^{-24}$ W or 0.3 ct/s in terms of event rate~\cite{HAYSTAC2021}. The technology used in the current generation of haloscope experiments can access axions with masses $m_a\sim 0.25-25\ \upmu$eV, but higher masses in the range $m_a\sim 40-200\:\ \upmu$eV, corresponding to photons at $\sim 10-50$ GHz, are favored for post-inflationary QCD axions~\cite{Borsanyi2016, Ballesteros2017, Co2020, Buschmann2022}. One challenge that must be addressed to perform haloscope searches at higher frequencies is the loss of signal. In order for signal to accumulate in the cavity, the resonant frequency $\nu_c$ must match the axion's mass. This leads to a reduction in conversion power at higher frequencies, as the cavity's radius and therefore volume decrease to maintain the resonant condition. In order to maintain sensitivity to QCD axions, haloscopes must either increase the signal by improving the axion-photon conversion rate or lower the noise to achieve sensitivity to smaller photon signals. Several such ideas have been proposed for next-generation experiments~\cite{Snowmass2022}.

The use of single-photon detectors for readout in next-generation haloscope experiments provides a promising way of addressing this challenge. Unlike linear amplifiers, single-photon detectors are not sensitive to the phase of a microwave field, allowing for measurements of the field which are not subject to the SQL and are instead limited by shot noise from thermal photons~\cite{Caves1982}. This thermal photon noise is frequency-dependent and at high frequencies $\gtrsim$ 10 GHz decreases enough that single-photon detectors will have lower noise power than linear amplifiers under reasonable assumptions~\cite{Lamoreaux2013}. This leads to an overall higher signal-to-noise ratio, enhancing the haloscope's ability to detect small signals. With the advancements in the field of microwave photon detection brought on by quantum computing~\cite{Schuster2007, Romero2009, Chen2011, Gu2017, Lescanne2020, Wang2021}, there are a number of potential schemes in development for single-photon detection in axion experiments~\cite{Dixit2021,Pankratov2022}.

In this work, we propose a design for a Rydberg atom-based single-photon detector which can be coupled to an axion haloscope to search for high-frequency axions. Rydberg atoms, with extremely high principal quantum number $n$, have many transitions in the $10-1000$ GHz range with large dipole transition moments making them well suited for single-photon detection at microwave frequencies. A straightforward detection scheme utilizing a transition between two neighboring Rydberg levels has been demonstrated to be sensitive to very small microwave fields around 14 GHz~\cite{Sedlacek2012}.

We begin in Sec.~\ref{sec: overview} with an overview of our single-photon detector concept and the advantages of single-photon detection for haloscope experiments in general. In Sec.~\ref{sec: rydberg}, we discuss which Rydberg states are appropriate for single-photon detection and show that photons in the frequency range $10-50$ GHz can readily be detected with Rydberg atoms. In Sec.~\ref{sec: design}, we propose a design for incorporating these Rydberg atoms into a haloscope experiment and estimate its efficiency and noise rate. Finally, in Sec.~\ref{sec: sensitivity}, we estimate the axion discovery sensitivity of our single-photon detector when connected to several different haloscope designs and show several designs that feature sensitivity to QCD axions with masses above 40 $\upmu$eV.

\section{Overview}\label{sec: overview}
\subsection{Experimental Concept Overview}\label{sec: concept}

In this paper, we propose a single-photon detection scheme based on Rydberg atoms for haloscope-type axion dark matter experiments. Rydberg atoms have previously been considered for axion searches in a proposed experiment with $^{85}$Rb Rydberg atoms in a haloscope tuned to 2.5 GHz~\cite{tada1999, Yamamoto2001}. Single-photon detection with Rydberg atoms, although applicable to a broad range of frequencies, is best suited to detecting photons at frequencies above 10 GHz~\cite{Lamoreaux2013}. At these higher frequencies, traditional sensors used in haloscope experiments are limited by quantum measurement noise and are less readily available.

\begin{figure}[h] \centering
\includegraphics[width=\columnwidth]{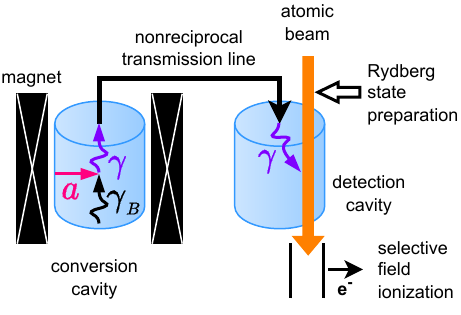}
\caption{Single-photon detection scheme for a haloscope experiment. Axions ($a$) convert into photons ($\gamma$) inside a conversion cavity by absorbing virtual photons from a strong magnetic field ($\gamma_B$). The axion-converted photon is coupled out of the conversion cavity and into the detection cavity via a non-reciprocal transmission line. Both of the cavities and the transmission line are cooled to sub-Kelvin temperatures by a dilution refrigerator (not shown). In the detection cavity, the photon couples to a beam of Rydberg atoms. The final state of the Rydberg atoms is read out using selective field ionization. \label{fig: bigpicture}}
\end{figure}

A schematic of the single-photon detection scheme incorporated into a haloscope experiment is shown in Fig.~\ref{fig: bigpicture}. As in standard cavity haloscope experiments, axion-photon conversion takes place inside a ``conversion cavity'': a microwave cavity placed inside a large magnetic field. Axions interact with the magnetic field and are converted into photons with frequencies $\nu_a \approx m_a c^2/h$. Photons resonant with the conversion cavity are collected, amplified, and coupled out of the cavity. A non-reciprocal transmission line sends the photons to a second cavity, the ``detection cavity'', which houses the single-photon detector away from the conversion cavity's magnetic field. Both cavities, along with the transmission line, are cooled to sub-Kelvin temperatures with a dilution refrigerator to reduce backgrounds from thermal photons. Single-photon detection takes place in the detection cavity, where photons couple to a beam of Rydberg atoms sent through the cavity. This coupling can take the form of absorption or stimulated emission, corresponding to an increase or decrease in the energy of the Rydberg atom's state. This change in the Rydberg state is detectable by selective field ionization (SFI), where the Rydberg atoms are placed in an electric field strong enough to ionize only one of the two states and an electron detector is used to see whether ionization has occurred. This combination of atom-photon coupling and state readout forms the single-photon detector.

The amount of signal power coupled out of the conversion cavity, $P_\text{sig}$, is given by the Sikivie equation:
\begin{equation}
    P_\text{sig} = \left(\frac{2 \alpha^2 \hbar^3  c^3}{\pi \mu_0}\frac{g_{\gamma}^2\rho_a}{\Lambda^4}\right) \left(B_0^2 V C_{mnl} \nu_c Q_\text{conv} \frac{\beta}{1+\beta}\right)\label{eq: sigpower}
\end{equation}

\noindent in the case where the cavity's resonant frequency exactly matches the axion frequency~\cite{Sikivie1983, Sikivie1985}. Physical constants and properties of dark matter appear in the left set of terms: $g_{\gamma}$ is the axion-photon coupling, $\rho_a = 0.45$ GeV/cm$^3$ is the local dark matter density~\cite{Read2014},  and $\Lambda = 77.6$ MeV describes the relation between the axion mass and the axion decay constant~\cite{Patrignani2016}. Experimental parameters appear in the right set of terms: $B_0$ is the magnetic field strength, $V$ is the conversion cavity volume, $C_{mnl}$ is the cavity form factor for the mode specified by $mnl$, $\nu_c$ is the resonant frequency of the conversion cavity, $Q_\text{conv}$ is the conversion cavity's loaded quality factor, and $\beta$ parameterizes the coupling of the cavity and relates the unloaded quality factor $Q_\text{conv}^{0}$ to the loaded quality factor as  $Q_\text{conv} ^{0} = Q_\text{conv}(1+\beta)$.

The signal for the single-photon detector is the number of axion-converted photons present in the detection cavity with frequencies inside the cavity's bandwidth. The number of signal counts seen by the single-photon detector in a time $\tau$ is related to the signal power $P_\text{sig}$ and signal count rate $R_\text{sig}$ by
\begin{equation}
    N_\text{sig,sp} = \eta_\text{sp} \frac{P_\text{sig} \tau}{h \nu_a} = \eta_\text{sp} R_\text{sig} \tau,
\end{equation}
\noindent where $\eta_\text{sp}$ accounts for all efficiency losses after the photon signal is coupled out of the conversion cavity.

Noise in the single-photon detector takes the form of additional counts in the detector $N_\text{sys,sp}$. Assuming that the statistical fluctuations in the number of signal and noise counts are equal to $\sqrt{N}$, the signal-to-noise ratio (SNR) for the single-photon detector is
\begin{equation}
    \text{SNR}_\text{sp} = \frac{N_\text{sig,sp}}{\sqrt{N_\text{sig,sp}+N_\text{sys,sp}}} = \frac{R_\text{sig,sp}}{\sqrt{R_\text{sig,sp}+R_\text{sys,sp}}} \sqrt{\tau}. \label{eq: snrsp}
\end{equation}

Alternately, Eq.~\ref{eq: snrsp} can be solved for $\tau$ to give the integration time needed to reach a desired target SNR, which we denote as $\tau_\text{SNR,sp}$. 

For haloscope axion searches, the standard figure of merit is the scan rate. The scan rate is defined as the time needed to perform an axion search across a given frequency range, reaching a target SNR for every frequency in that range. This incorporates information about the haloscope's frequency tuning that is not captured by the ratio of SNRs at a single frequency. 

A haloscope search using a single-photon detector collects counts within the detector bandwidth at each tuning step. The counts collected from this tuning step give sensitivity to axion signals appearing in the entire detection cavity bandwidth $\Delta \nu_\text{det}$, but cannot resolve individual frequencies within the bandwidth. Ideally, then, $\Delta \nu_\text{det}$ should be matched to the axion signal's bandwidth $\Delta \nu_a$ to avoid excess noise pickup outside the signal bandwidth while also not attenuating the signal. If $\Delta \nu_\text{det} \approx \Delta \nu_a$, the scan rate is given by
\begin{equation}
     \left(\frac{d\nu}{dt}\right)_\text{sp} = \frac{\Delta \nu_\text{det}}{\tau_\text{SNR,sp}} = \frac{\Delta \nu_\text{det}}{\text{SNR}^2}\frac{R_\text{sig,sp}^2}{R_\text{sig,sp}+R_\text{sys,sp}} \label{eq: scan_sp}.
\end{equation}

The noise in a single-photon detector, $R_\text{sys,sp}$, is fundamentally driven by thermal photons produced from blackbody radiation (BBR), which produce identical signals to axion-converted photons when inside the detector's bandwidth. However, the number of thermal photons drops exponentially at higher frequencies and lower operating temperatures. For frequencies on the order of 10 GHz and temperatures on the order of 10 mK, the thermal noise background is negligible, leading to high scan rates unachievable with a standard linear amplifier readout~\cite{Lamoreaux2013}.

\subsection{Advantages of Single-Photon Detection}\label{sec: advantage}
Single-photon detection is best suited for high frequencies above 10 GHz, where haloscope experiments must contend with decreased signal power due to shrinking cavity size. Resonance between the conversion cavity and the axion-converted photons requires that the cavity's radius scale with the photon wavelength, while the cavity's length should scale with the cavity's radius to avoid introducing mode crossings into the cavity's modes~\cite{Bai2023}. Then, $V \propto \nu_a^{-3}$, and the overall signal rate goes as
\begin{equation}
    R_\text{sig} \propto m_a^{-3},\label{eq: scaling}
\end{equation}

\noindent slowing the scan rate of any haloscope searching for axions at higher masses. However, with single-photon detection available at higher frequencies, dramatic enhancements in scan rate become possible, counteracting this drop in signal.

The scan rate of a haloscope using a linear amplifier for readout can be written in the same notation as Eq.~\ref{eq: scan_sp}:
\begin{equation}
\left(\frac{d\nu}{dt}\right)_\text{la} = \frac{4}{5} \frac{\Delta \nu_\text{conv}}{\text{SNR}^2} \frac{R_\text{sig,la}^2}{N_\text{sys,la}^2 \Delta \nu_a}\label{eq: scan_la}
\end{equation}

\noindent\cite{Alkenany2017}. The overall form of the single-photon and linear-amplifier scan rates is similar. As we will see, the single-photon enhancement is primarily driven by the difference in the noise rates between the two types of readout.

The noise rate of a linear-amplifier-based haloscope is fundamentally limited by quantum measurement noise. The lower bound on the quantum measurement noise, known as the standard quantum limit (SQL), is $N_\text{sys,la} = 1$, independent of frequency~\cite{Haus1962, Caves1982}. Haloscope experiments have reached SQL-limited noise levels and even used techniques such as squeezed states to achieve measurement noise 4 dB below the SQL at the cost of sacrificing some of their signal~\cite{HAYSTAC2021}. Single-photon detection avoids measurement noise entirely due to its phase insensitivity, which allows measurements in a basis where the signal amplitude is free of noise without needing to discard any signal power.

Because single-photon detectors can eliminate measurement noise, they are able to more quickly scan for axions across broad frequency ranges, making them ideal for experiments aiming to discover an axion signal at an unknown frequency. While the axion mass is unknown, scan rate is the most important consideration. However, if an axion signal were discovered, linear amplifiers would have an important role to play due to their frequency resolution: although vulnerable to additional sources of noise, linear amplifiers can provide insight into the axion's properties through resolution of the axion lineshape~\cite{Zheng2016}. For this reason, linear amplifier technology should remain a key part of haloscope development. 

The remarks in this section are general to all forms of single-photon detection in haloscope experiments. However, in the rest of this paper, we focus on the Rydberg-atom-based single-photon detection scheme outlined in \ref{sec: concept}, showing how it is capable of achieving the performance described here.

\section{Rydberg Atoms for Single-Photon Detection}\label{sec: rydberg}

The single-photon detection scheme outlined in the previous section uses Rydberg atoms as detectors for microwave photons created from axion-photon conversion. Many transitions between Rydberg states are at microwave frequencies and take place quickly due to the high transition dipole moments of the states, making Rydberg atoms a compelling single-photon detector for this application. Here, we describe the design considerations that inform the choice of which Rydberg transitions to use for single-photon detection and show that axion-converted photons with frequencies from 10 GHz to 50 GHz can be detected with our Rydberg atom design.

\subsection{Choice of Atomic Species}
The first choice that must be made in the design of the Rydberg-atom-based single-photon detector is which atom to use. Alkali atoms are relatively easy to work with, benefiting from well-established techniques for state preparation and spectroscopy. From an ease-of-use standpoint, rubidium atoms are particularly compelling. However, previous attempts to use $^{85}$Rb Rydberg atoms for photon detection by the CARRACK collaboration found that the efficiency of atom-photon coupling was very sensitive to stray electric fields inside the detection region \cite{TADA2006, Haseyama2008}. Based on this, CARRACK estimated the size of the stray electric field effect for different species of Rydberg atoms and proposed that using $^{39}$K could increase the viability of single-photon detection \cite{Haseyama2008}. 

We further investigate this effect, comparing the relative susceptibility to stray electric fields for several Rydberg states in $^{85}$Rb and $^{39}$K with different principal quantum numbers $n$. First, we estimate each Rydberg state's energy dependence on the electric field $\mathcal E$ at low fields via its scalar polarizability $\alpha_0$, where $\Delta E = \frac{1}{2}\alpha_0 \mathcal E ^2$. Then, we use the alkali Rydberg calculator~\cite{ARC} (ARC) to calculate the scalar polarizability for $nS$ and $nP_{1/2}$ states in $^{85}$Rb and $^{39}$K. The scalar polarizability has an approximate $n$ dependence of $\alpha_0 \propto n^7$, which is consistent with our ARC calculations and allows us to interpolate across arbitrary values of $n$~\cite{gallagher1994}. We define $\Delta \alpha_0$ as the difference between the scalar polarizabilities of the $nS$ and $nP$ states. When $\Delta \alpha_0$ is small, the transition energy is robust to stray electric fields regardless of the value of $\alpha_0$: the energy of each state might change, but the transition energy stays the same.
 
\begin{figure}\centering
    \includegraphics[width=\linewidth]{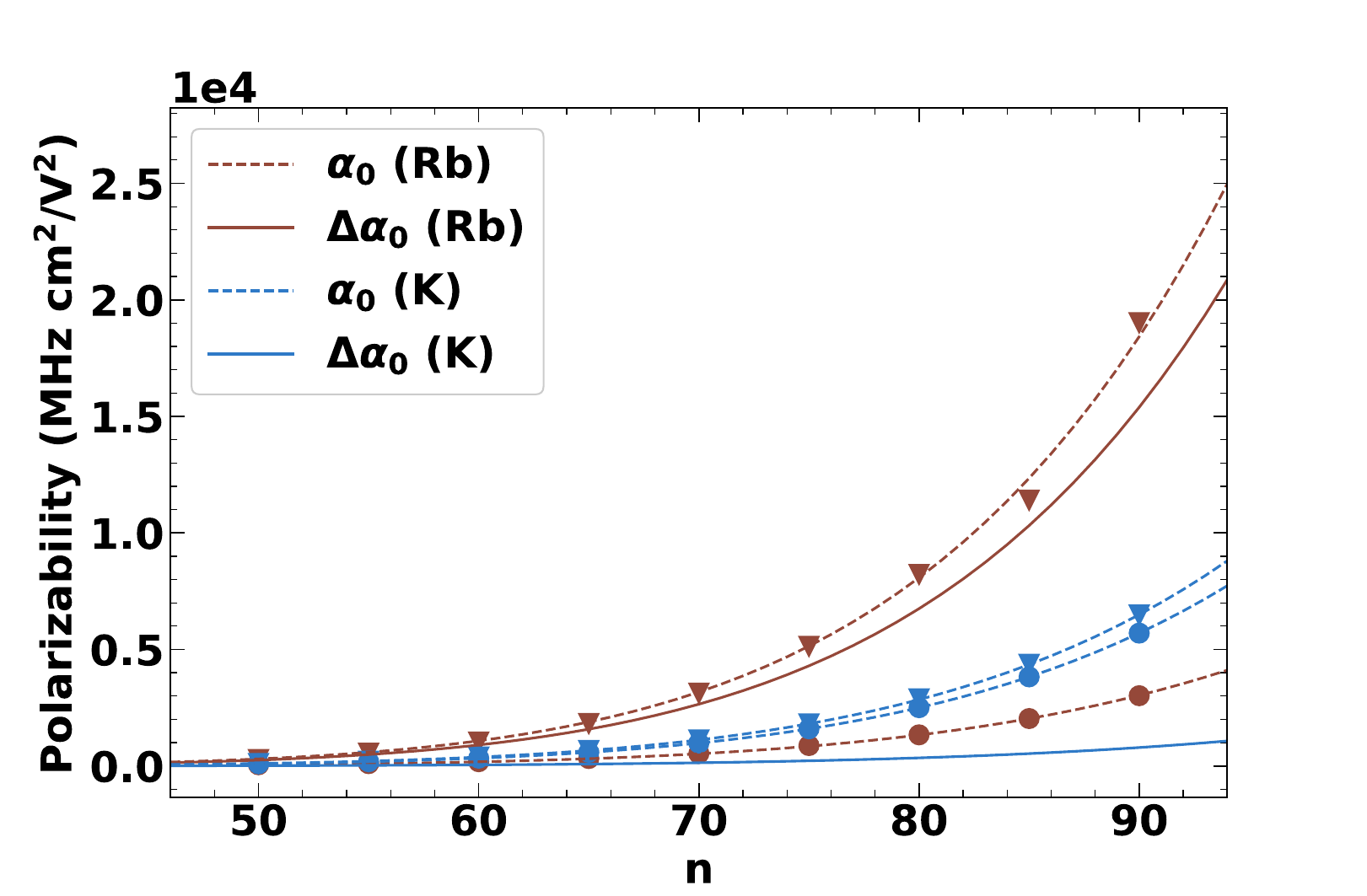}
    \caption{Scalar polarizability $\alpha_0$ (calculated with ARC) vs. $n$ for $nS$ and $nP_{1/2}$ Rydberg states in $^{85}$Rb and $^{39}$K. Dashed lines indicate fits to the approximate formula $\alpha_0 \propto n^7$ for $nS$ and $nP_{1/2}$ states, marked with circles and triangles respectively. Solid lines indicate $\Delta \alpha_0$.
    }
    \label{fig: pol}
\end{figure}

 In Fig.~\ref{fig: pol}, we plot the values of $\alpha_0$ and $\Delta \alpha_0$ for $^{85}$Rb and $^{39}$K as a function of $n$. While the overall polarizability increases with $n$ for both types of atom, the polarizability of the $nS$ and $nP$ states stays close together for $^{39}$K and diverges rapidly for $^{85}$Rb, leading to a large difference in $\Delta \alpha_0$. This indicates that $^{39}$K is well-suited for avoiding the effects of stray electric fields, especially for higher-$n$ states. For this reason, we propose to use $^{39}$K Rydberg atoms for our Rydberg-atom-based single-photon detector.

\subsection{Choice of Rydberg States}\label{sec: states}

\begin{figure*} \centering
\begin{subfigure}{.285\textwidth}\centering\caption{}    \label{fig: statepreplevels} 
\resizebox{.8\textwidth}{!}{
\begin{tikzpicture}[font=\bf\sffamily, scale=1.4, level/.style={thick}, trans/.style={thick,<->, >=stealth,shorten >=7pt, shorten <=7pt}, laser/.style={ultra thick, <->, >=stealth, shorten >=0.5, shorten <=0.5, red}, p/.style = {midway, fill=white, inner sep=1}
  ]
\pgfmathsetmacro\s{-.5} 
\pgfmathsetmacro\p{3} 
\pgfmathsetmacro\ss{4.5} 
 \pgfmathsetmacro\h{.15} 
\pgfmathsetmacro\g{-.3} 
\pgfmathsetmacro\e{2} 
\pgfmathsetmacro\w{2} 
\definecolor{bluelight}{rgb}{.51, 0, .78}
\definecolor{orange2}{rgb}{1, .31 0}	

    \node at (\s-.2,\g/2+\h/2){$\bm{\mathsf{4^2S_{1/2}}}$};
\node at (\s,\p){$\bm{\mathsf{5^2P_{1/2}}}$};
\node at (\s, \ss){$\bm{\mathsf{n^2S_{1/2}}}$};

\draw[level] (0,\h)--++(1,0) node[right]{$\bm{\mathsf{F=2}}$} (0,\g)--++(1,0)node[right]{$\bm{\mathsf{F=1}}$}  (0,\p)--++(1,0);
 \draw[thick,dotted] (0,\g)--(\s+.2,\g/2+\h/2)--(0,\h);
\draw[trans](\s,\p)--(\s,\g/2+\h/2);
\node[fill=white,inner sep=.1] at (\s, \p/2){\specialcell{404.8 nm\\ $\bm{\mathsf{\approx}}$}}; 
\draw[level]  (0,\ss)--(1,\ss);
\draw[trans](\s,\ss)--(\s,\p) node[midway,fill=white,inner sep=.1]{\specialcell{$\bm{\mathsf{\sim}}$ 970 nm\\ $\bm{\mathsf{\approx}}$}};

\draw[laser,bluelight] (.5,\h)--(.5,\p) node[p] {$\bm{\mathsf{P}}$};
\draw[laser, orange2] (.5,\p)--(.5,\ss) node[p] {$\bm{ \mathsf{S}}$};

\end{tikzpicture}
}
\end{subfigure}%
\begin{subfigure}{.3\textwidth}\centering\caption{}    \label{fig: abstrans} 
\resizebox{\textwidth}{!}{
\begin{tikzpicture}[font=\bf\sffamily\large, scale=2, level/.style={thick}, level2/.style={ultra thick}, connect/.style={thick},trans/.style={thick,<->, >=stealth,shorten >=7pt, shorten <=7pt}, laser/.style={ultra thick, <->, >=stealth, shorten >=0.5, shorten <=0.5, red}, p/.style = {midway, fill=white, inner sep=1}
  ]
\pgfmathsetmacro\s{-.5} 
\pgfmathsetmacro\p{3} 
\pgfmathsetmacro\ss{4.5} 
 \pgfmathsetmacro\h{.15} 
\pgfmathsetmacro\g{-.3} 
\pgfmathsetmacro\e{2} 
\pgfmathsetmacro\w{2} 
\definecolor{bluelight}{rgb}{.51, 0, .78}
\definecolor{orange2}{rgb}{1, .31 0}	
 

\node at (-0.5, 2) {$\bm{\mathsf{(n)}}$};
\draw[level] (-0.25,2) -- (0.15,2);
\draw[connect] (0.15,2) -- (0.65,5.75/2);
\draw[level] (0.65,5.75/2) -- (1.75,5.75/2) node[right] {$\bm{\mathsf{F_{5/2}}}$};
\draw[connect] (0.15,2) -- (0.75,4.75/2);
\draw[level2] (0.75,4.75/2) -- (1.75,4.75/2) node[right] {$\bm{\mathsf{D_{3/2}}}$};
\draw[connect] (0.15,2) -- (0.5,3.5/2);
\draw[level] (0.65,3.5/2) -- (0.5,3.5/2); 
\draw[connect] (0.65,3.5/2) -- (0.9,3.75/2) (0.65,3.5/2) -- (0.9,3/2);
\draw[level] (0.9,3.75/2) -- (1.75,3.75/2) node[right] {$\bm{\mathsf{P_{3/2}}}$};
\draw[level] (0.9,3/2) -- (1.75,3/2) node[right] {$\bm{\mathsf{P_{1/2}}}$};
\draw[connect] (0.15,2) -- (0.65,1);
\draw[level2] (0.65,1) -- (1.75,1) node[right] {$\bm{\mathsf{S_{1/2}}}$};

\node at (-0.5, 4.5) {$\bm{\mathsf{(n+2)}}$};
\draw[level] (-0.05,2+2.5) -- (0.25,2+2.5);
\draw[connect] (0.25,4.5) -- (0.5,3.5/2+2.5) (0.25,2+2.5) -- (0.65,5);
\draw[level] (0.65,2.5+2.5) -- (1.75,2.5+2.5) node[right] {$\bm{\mathsf{F_{5/2}}}$};
\draw[level] (0.65,3.5/2+2.5) -- (0.5,3.5/2+2.5);
\draw[connect] (0.65,3.5/2+2.5) -- (0.9,3.75/2+2.5) (0.65,3.5/2+2.5) -- (0.9,3/2+2.5);
\draw[level] (0.9,3/2+2.5) -- (1.75,3/2+2.5) node[right] {$\bm{\mathsf{P_{1/2}}}$};
\draw[level] (0.9,3.75/2+2.5) -- (1.75,3.75/2+2.5) node[right] {$\bm{\mathsf{P_{3/2}}}$};

\draw[excite] (1.1,4.75/2) -- (1.1,3/2+2.5); 
\draw[excite] (1.4,4.75/2) -- (1.4,5.75/2); 
\draw[excite] (1.25,1) -- (1.25,3/2); 
\draw[excite] (1.55,1) -- (1.55,3.75/2); 

\end{tikzpicture}
}
\end{subfigure}%
\begin{subfigure}{.3\textwidth}\centering\caption{}    \label{fig: emtrans} 
\resizebox{\textwidth}{!}{
\begin{tikzpicture}[font=\bf\sffamily\large, scale=2, level/.style={thick}, level2/.style={ultra thick},connect/.style={thick},trans/.style={thick,<->, >=stealth,shorten >=7pt, shorten <=7pt}, laser/.style={ultra thick, <->, >=stealth, shorten >=0.5, shorten <=0.5, red}, p/.style = {midway, fill=white, inner sep=1}
  ]
\pgfmathsetmacro\s{-.5} 
\pgfmathsetmacro\p{3} 
\pgfmathsetmacro\ss{4.5} 
 \pgfmathsetmacro\h{.15} 
\pgfmathsetmacro\g{-.3} 
\pgfmathsetmacro\e{2} 
\pgfmathsetmacro\w{2} 
\definecolor{bluelight}{rgb}{.51, 0, .78}
\definecolor{orange2}{rgb}{1, .31 0}	
 

\node at (-0.5, 2.3) {$\bm{\mathsf{(n+1)}}$};
\draw[level] (-0.05,2.3) -- (1.75,2.3) node[right] {$\bm{\mathsf{P_{1/2}}}$};

\node at (-0.5, 2-0.5) {$\bm{\mathsf{(n)}}$};
\draw[level] (-0.25,2-0.5) -- (0.15,2-0.5);
\draw[connect] (0.15,2-0.5) -- (0.75,5.75/2);
\draw[level2] (0.75,5.75/2) -- (1.75,5.75/2) node[right] {$\bm{\mathsf{D_{3/2}}}$};
\draw[connect] (0.15,2-0.5) -- (0.5,3.5/2-0.5);
\draw[level] (0.65,3.5/2-0.5) -- (0.5,3.5/2-0.5); 
\draw[connect] (0.65,3.5/2-0.5) -- (0.9,3.75/2-0.5) (0.65,3.5/2-0.5) -- (0.9,3/2-0.5);
\draw[level] (0.9,3.75/2-0.5) -- (1.75,3.75/2-0.5) node[right] {$\bm{\mathsf{P_{3/2}}}$};
\draw[level] (0.9,3/2-0.5) -- (1.75,3/2-0.5) node[right] {$\bm{\mathsf{P_{1/2}}}$};
\draw[connect] (0.15,2-0.5) -- (0.65,1-0.5);
\draw[level2] (0.65,1-0.5) -- (1.75,1-0.5) node[right] {$\bm{\mathsf{S_{1/2}}}$};

\node at (-0.5, -0.625) {$\bm{\mathsf{(n-1)}}$};
\draw[level] (-0.05,-0.625) -- (0.25,-0.625);
 \draw[connect] (0.25,-0.625) -- (0.5,-0.125-0.75) (0.25,-0.625) -- (0.65,-0.125);
 \draw[level] (0.65,-0.125) -- (1.75,-0.125) node[right] {$\bm{\mathsf{F_{5/2}}}$};
 \draw[level] (0.65,-0.125-0.75) -- (0.5,-0.125-0.75);
 \draw[connect] (0.65,-0.125-0.75) -- (0.9,-0.125-1+0.375) (0.65,-0.125-0.75) -- (0.9,-0.125-1);
 \draw[level] (0.9,-0.125-1) -- (1.75,-0.125-1) node[right] {$\bm{\mathsf{P_{1/2}}}$};
 \draw[level] (0.9,-0.125-1+0.375) -- (1.75,-0.125-1+0.375) node[right] {$\bm{\mathsf{P_{3/2}}}$};

\draw[emit] (1.1,5.75/2) -- (1.1,-0.125); 
\draw[emit] (1.3,1-0.5) -- (1.3,-0.125-1+0.375); 
\draw[emit] (1.5,1-0.5) -- (1.5,-0.125-1); 
\draw[emit] (0.9,5.75/2) -- (0.9,2.3); 
\end{tikzpicture}
}
\end{subfigure}%
\caption{Partial energy level diagrams of $^{39}$K. (a) Energy level diagram showing two-photon excitation used to prepare the Rydberg state $n^2S_{1/2}$. This technique can be used to prepare either $nS_{1/2}$ or $nD_{3/2}$ states. (b) A subset of Rydberg transitions accessible via photon absorption from the prepared states. (c) A subset of Rydberg transitions accessible via photon emission from the prepared states. } \label{fig: states} 
\end{figure*}
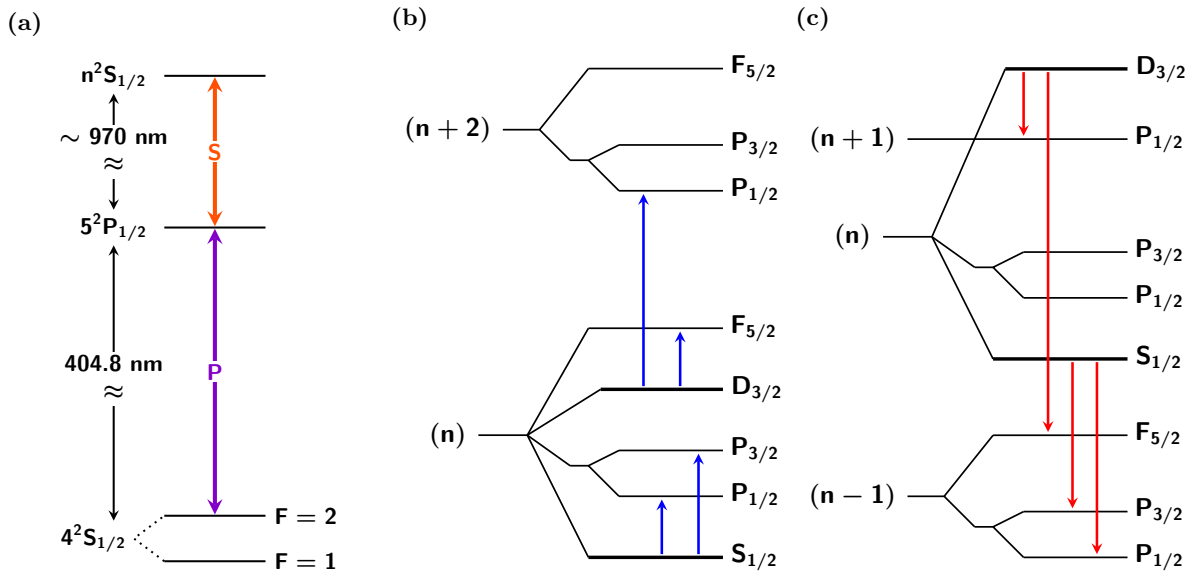

Rydberg states are prepared by exciting ground-state atoms into a high-$n$ state. We use a two-photon transition for state preparation, passing from the 4$S_{1/2}$ ground state to the 5$P_{1/2}$ intermediate state, and then from this intermediate state to an n$S_{1/2}$ or n$D_{3/2}$ Rydberg state. This is illustrated in Fig.~\ref{fig: statepreplevels} for $^{39}$K. Using this state preparation scheme, we have previously created and characterized over fifty different Rydberg states in $^{39}$K~\cite{Zhu2022}. 

From each starting state accessible with state preparation, many different Rydberg-Rydberg transitions are possible, driven by absorption or stimulated emission. Due to this variety of starting and ending states, there are multiple Rydberg transitions resonant with each desired photon frequency. Choosing which transitions to use is mainly motivated by two properties: transition dipole moment and state lifetime. The transition dipole moment $d$ determines how quickly the atom-photon interaction takes place. The detection cavity can only store photons for a short time determined by its quality factor, with typical storage times on the order of a few $\upmu$s. For a high probability of atom-photon interaction, then, a large transition dipole moment is necessary. A long state lifetime is also necessary to ensure that the atom will not decay out of its Rydberg state during the single-photon detector's operation. Rydberg states are long-lived with typical lifetimes on the order of 100 $\upmu$s, much longer than the cavity's photon storage time, so the lifetime constraints are primarily set by the time needed for the atoms to travel between the initial state preparation region and the final readout region. The exact requirements for $d$ and $\tau$ depend on the experimental design, with the dipole moment more heavily constrained due to the short time that photons spend in the detection cavity. 

Achieving suitable $d$ and $\tau$ is easier at higher operating frequencies. Transition frequency tends to increase with $n$, and both dipole moment and lifetime approximately scale with $n$ as $d \propto n^2$ and $\tau \propto n^3$ ~\cite{gallagher1994}. For a fixed transition frequency, we can optimize $d$ by looking at several different transitions around that frequency. To find the transitions with the strongest dipole moments, we calculate $d$ using ARC for all of the allowed transitions from the states that can be created with our state preparation scheme. We then select the transitions with the highest $d$ for further investigation.

We show the properties of the allowed transitions with the highest dipole moments in Table~\ref{tab: states}, choosing transitions with frequencies of 16$\pm 0.5$ GHz, near the bottom of our proposed frequency range. For each transition, the starting state is listed followed by the ending state. The dipole moment $d$ depends on the difference between the total angular momentum quantum numbers $m_j$ of the starting and ending states. For this table, we focus on $\Delta m_j=\pm1$ which has the largest dipole moment and therefore the largest transition probability. The Rydberg state lifetimes are listed for the starting states ($\tau_n$) and ending states ($\tau_{n'}$). The absorption transitions in the table are illustrated in Fig.~\ref{fig: abstrans} and the emission transitions in the table are illustrated in Fig.~\ref{fig: emtrans}.

\begin{table}[h]\centering
\caption{Rydberg transition frequency, dipole moment, and lifetimes of the starting and ending states near 16 GHz.} \label{tab: states}
\begin{tabular}{c c c c c} \hline\hline
Transition  & $\nu$ (GHz) & $d\ (ea_0)$ & $\tau_n\ (\upmu$s) & $\tau_{n'}\ (\upmu$s)\\
  \hline
60$S_{1/2}\rightarrow$ 60$P_{3/2}$ & 15.78 & 2251 & 227 & 781\\
\hline
63$S_{1/2}\rightarrow$ 62$P_{3/2}$ & 15.73 & 2230 & 265 & 841\\
\hline
57$D_{3/2}\rightarrow$ 58$P_{1/2}$ & 15.94 & 2225 & 453 & 704\\
\hline
61$D_{3/2}\rightarrow$ 63$P_{1/2}$ & 16.31 & 2141 & 555 & 909\\
\hline
68$D_{3/2}\rightarrow$ 67$F_{5/2}$ & 15.79 & 2062 & 768 & 228\\
\hline
48$D_{3/2}\rightarrow$ 48$F_{5/2}$ & 16.00 & 2024 & 271 & 84\\
\hline
60$S_{1/2}\rightarrow$ 60$P_{1/2}$ & 15.68 & 1845 & 227 & 781\\
\hline
63$S_{1/2}\rightarrow$ 62$P_{1/2}$ & 15.81 & 1813 & 265 & 865\\

\hline\hline
\end{tabular}
\end{table} 

\subsection{Rydberg Frequency Tuning}

\begin{figure*}[htb]\centering
    \includegraphics[width=\linewidth]{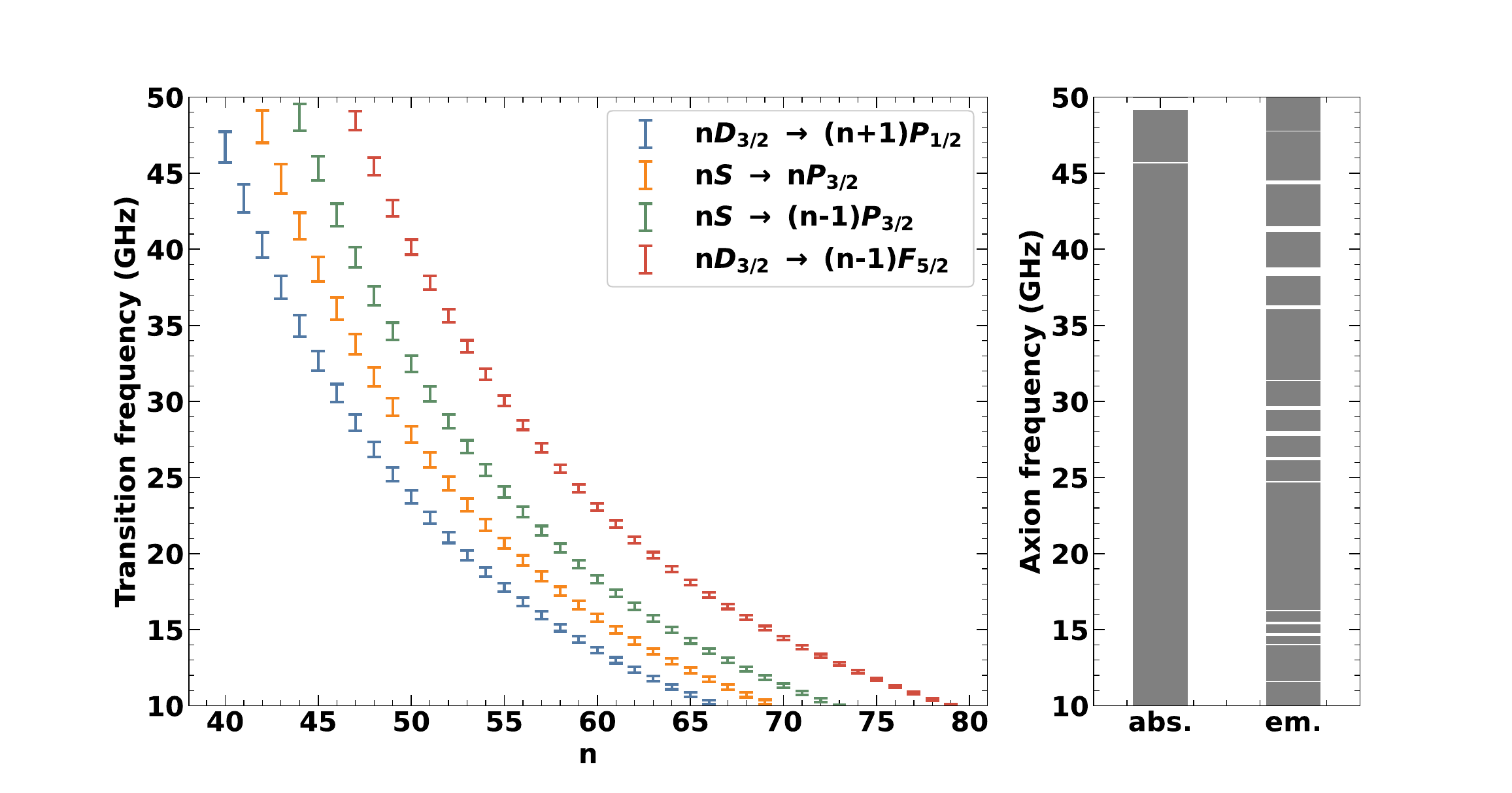}
    \caption{
    Left: Zeeman-tuned transition frequency $\nu_\text{Ry}+\Delta \nu_\text{Zeeman}$ vs. $n$ for the transitions $nD_{3/2}\ \rightarrow\ (n+1)P_{1/2}$, $nS\ \rightarrow\ nP_{3/2}$, $nS\ \rightarrow\ (n-1)P_{3/2}$, and $nD_{3/2}\ \rightarrow\ (n-1)F_{5/2}$. The vertical bars indicate the maximum Zeeman-tuning range of the individual transitions. Right: Range of accessible $\nu_a$ for all absorption transitions (abs.) in Fig.~\ref{fig: abstrans} and emission transitions (em.) in Fig.~\ref{fig: emtrans}.
    }
    \label{fig: tuning}
\end{figure*}

Performing an axion search using Rydberg-atom-based single-photon detectors will require tuning the single-photon detector to search for axions at many different frequencies. We can broadly ``tune'' the detection frequency of the Rydberg atoms by choosing different transitions, making large changes in frequency by increasing $n$ and smaller changes in frequency by using a different transition type at a similar frequency.  However, an additional tuning mechanism is necessary to fill out the intermediate frequencies between these resonances. 

Rydberg transition frequencies can be shifted by either external electric fields (the Stark effect) or external magnetic fields (the Zeeman effect). We propose to use Zeeman shifts as our fine-tuning mechanism, since $^{39}$K transition frequencies are relatively robust to external electric fields. For magnetic fields up to a few hundred Gauss, Zeeman tuning produces linear energy shifts that are easy to control. The frequency range of Zeeman tuning is limited by the magnetic field level where higher-order effects begin to dominate, which scales with $n$ as $1/n^4$~\cite{gallagher1994}. 

Fig.~\ref{fig: tuning} shows the axion frequencies $\nu_a$ that can be reached in the $10-50$ GHz range  (or $\sim 40-200\:\upmu$eV in mass units), calculated based on the transitions discussed in \ref{sec: states} and modified with linear Zeeman tuning. Through Zeeman tuning of absorption transitions, ~98\% of this mass range is readily accessible, and through Zeeman tuning of emission transitions, ~92\% of this mass range is readily accessible. Using Zeeman tuning of both absorption and emission transitions, the entire mass range is accessible.

\section{Single Photon Detector Design}\label{sec: design}
We now examine the overall single-photon detection scheme incorporating the Rydberg states discussed in Sec.~\ref{sec: rydberg}. In Sec.~\ref{sec: coupling} and Sec.~\ref{sec: beamsource}, we discuss the design considerations informing the choice of detection cavity, atomic beam source, and state readout scheme. Based on this design, we estimate the efficiency and noise rate of the single-photon detector in Sec.~\ref{sec: eff} and Sec.~\ref{sec: noise}.

\subsection{Detection Cavity Quality Factor \label{sec: coupling}}
The quality factor of the detection cavity, $Q_\text{det}$, determines how long photon signals will persist in the cavity before decaying. For efficient single-photon detection, $Q_\text{det}$ should be chosen so that the detector operates in the strong coupling regime, where the overall coupling rate $g_\text{dipole}$ between Rydberg atoms and the microwave field is greater than any decay rates~\cite{Thompson1992}.

The strong coupling conditions require $g_\text{dipole}>1/\tau_\text{Ry}, 1/\tau_{\gamma}$, where $\tau_\text{Ry}$ and $\tau_{\gamma}$ are the lifetimes of the atom and the photon respectively. The photon lifetime depends on the detection cavity's loaded quality factor $Q_\text{det}$ and its resonant frequency $\nu$: 
\begin{equation}
    \tau_\gamma = Q_\text{det}/(2\pi\nu).
\end{equation} 

For single-photon detection in an axion experiment, the ideal $Q$ is matched to the axion's quality factor $Q_a \sim \nu_a/\Delta\nu_a \sim 10^6$, where $\Delta \nu_a$ is the width of the axion's spectral distribution~\cite{Lamoreaux2013}. As discussed in Sec.~\ref{sec: states}, $\tau_{\gamma} \ll \tau_{\text{Ry}}$, so the limiting requirement of the strong coupling condition is 
\begin{equation}
    g_\text{dipole} > \frac{2\pi \nu_a}{Q_\text{det}}.\label{eq: strongcoup}
\end{equation}

It follows that the minimum $g_\text{dipole}$ required for strong atom-microwave coupling ranges from $g_\text{dipole}^\text{min}\sim \pi\cdot(20$ kHz) for 10 GHz photons to $g_\text{dipole}^\text{min}\sim \pi\cdot(100$ kHz) for 50 GHz photons. We can determine the coupling between a single atom and a single photon using the relation
\begin{equation}
    g_\text{dipole} = \frac{2d \mathcal{E}_\text{rms}}{\pi \hbar} \label{eq: gdip}
\end{equation}

\noindent~\cite{Raimond2001}. Here, $d$ is the dipole moment of the Rydberg transition induced in the atom by the photon and $\mathcal{E}_\text{rms}$ is the rms electric field generated by the photon.

We can estimate $\mathcal{E}_\text{rms}$ by assuming a specific cavity geometry. Suppose our cavity size is comparable to the microwave wavelength $\lambda_a=c/\nu_a$ and its volume is $V\sim (\lambda_a/2)^3$. The rms field per photon is then
\begin{equation}
    \mathcal E_\text{rms} \sim \sqrt{\frac{h\nu_a}{\epsilon_0 V}} \propto \nu_a^2 \label{eq: field}
\end{equation}

\noindent where $\epsilon_0$ denotes vacuum permittivity. Typical values of $\mathcal E_\text{rms}$ range from $\sim 5\upmu$V/cm at 10 GHz to $\sim 117\upmu$V/cm at 50 GHz. 

Combining Eqs.~\ref{eq: strongcoup} and \ref{eq: gdip}, we can rewrite the strong coupling condition in terms of $d$ and $\mathcal{E}_\text{rms}$ as
\begin{equation}
    \frac{d \mathcal{E}_\text{rms}}{\hbar} > \frac{\pi^2 \nu_a}{Q_\text{det}} \label{eq: strongcoup2}
\end{equation}

\noindent and use this and Eq.~\ref{eq: field} to derive a lower bound on $d$ as a function of $\nu_a$ and $Q_\text{det}$:
\begin{equation}
    d > \frac{\hbar \pi^2 \nu_a}{\mathcal{E}_\text{rms} Q_\text{det}} \propto \frac{1}{Q_\text{det} \nu_a}. \label{eq: strongcoup3}
\end{equation}

Thus, the minimum $d$ required for strong coupling is inversely proportional to $\nu_a$ and $Q_\text{det}$. With $\nu_a \sim 16$ GHz and $Q_\text{det} \sim 10^6$, this bound is $d > 1600 ea_0$, which is satisfied by all of the transitions given in Table~\ref{tab: states}. At lower frequencies, the resonant states have larger $n$ and therefore larger $d$. At higher frequencies, the lower bound on $d$ is less strict. Across our proposed frequency range, $Q_\text{det} \sim 10^6$ is sufficient for strong coupling.

However, single-photon detection can still take place with lower values of $Q_\text{det}$. Increasing the number of Rydberg atoms in the detection cavity, $N$, can increase the atom-photon coupling rate by a factor of $\sqrt{N}$ due to collective behavior of the Rydberg ensemble~\cite{haroche1985, Gross1979, GROSS1982theory}. Alternately, the strong coupling condition can be relaxed, corresponding to photon detection with a lower efficiency~\cite{Yamamoto1999}. For the remainder of this paper, we assume that strong coupling is achievable.

\subsection{Atom-Cavity Interaction Time \label{sec: beamsource}}
Once the Rydberg atom has coupled to a photon, the state of the atom-photon system undergoes Rabi oscillations. For effective single-photon detection, we must control the time the atom spends in the cavity relative to the frequency of these oscillations to maximize the probability that the atom undergoes a change in state. This is primarily accomplished by controlling the speed of the atomic beam passing through the cavity.

The longitudinal velocity of the Rydberg atoms $v_L$ is broadly controlled by the choice of beam source and can be further fine-tuned during the Rydberg state preparation step. Velocity-selective state preparation can be achieved by pumping only specific velocity classes of atoms into the ground state of the Rydberg excitation scheme~\cite{Main2021}. In our experiment, $^{39}$K atoms at speed $v_L$ would be pumped to the F=2 level of 4$^2S_{1/2}$ and, as shown in Fig.~\ref{fig: statepreplevels},  undergo a two-photon transition to the desired Rydberg state.

In the strong coupling regime, the ideal transient time through the microwave region ($\tau_t = L/v_L$) is 
\begin{equation}
    \tau_t=\pi/g_\text{dipole},
\end{equation}

\noindent which corresponds to a single $\pi$ pulse applied to the atom and photon's initial state. 

The strong coupling condition (Eq.~\ref{eq: strongcoup}) therefore gives an upper bound for $\tau_t$ and a lower bound for $v_L$. For the cubic cavity geometry assumed in the previous section, $L\sim \lambda_a/2$, giving
\begin{equation}
    v_L > \frac{2 \nu_a L}{Q_\text{det}} \sim \frac{c}{Q_\text{det}} \sim 300\ \mathrm{ m/s}. \label{eq: vbound}
\end{equation}

This constraint on the transverse velocity informs the choice of an appropriate beam source for this experiment. While the Rydberg state preparation would take place in the cryostat as close to the detection cavity as possible to minimize unwanted interactions and decays, the creation of the atomic beam itself would take place at room temperature. In Table~\ref{tab: beamsource}, we list typical values of $v_L$ and beam flux for three common types of alkali atom beam source. The $v_L$ estimated above most closely corresponds to the typical $v_L$ of a thermal beam. All three beam sources are capable of producing a high flux of alkali atoms.

\begin{table}[h]\centering
\caption{Types of beam sources, their typical velocities, and fluxes for alkali atoms.} \label{tab: beamsource}
\begin{tabular}{c c c} \hline\hline
Type  & $v_L$ & Flux\\
  \hline
thermal beam~\cite{Wei2022} &  $100-500$ m/s & $10^{11}$ atoms/s \\
2D MOT beam~\cite{Catani2006} &  $28-35$ m/s & $10^{11}$ atoms/s\\
supersonic beam~\cite{Larsen1974} & $1-5$ km/s & $10^{16}$atoms/s/sr\\
\hline\hline
\end{tabular}
\end{table} 

Different designs can make other beam sources more favorable than a thermal beam. The above estimate assumes that the strong coupling condition is barely achieved with $g_\text{dipole} \sim g_\text{min.}$. If the coupling rate is faster, then the ideal interaction time will be lower. Similarly, for a lower value of $Q_\text{det}$, the ideal interaction time will also be lower to match the lowered photon lifetime. For modest changes in $g_\text{dipole}$ or $Q_\text{det}$ from the above values, a supersonic beam would be more appropriate. However, for $Q_\text{det} \lesssim 10^5$ a beam with the $v_L$ required for the desired atom-photon interaction would be challenging to produce and control.

\subsection{Photon detector efficiency \label{sec: eff}}
We now turn our attention to the overall efficiency of the Rydberg-atom-based single-photon detector. We introduce an overall photon detection efficiency $\eta$ to account for all losses taking place after the axion-converted photon is coupled out of the conversion cavity.

The conversion cavity is connected to the detection cavity via a shared transmission line with a circulator in between. We add this circulator to ensure that signals only flow from the conversion cavity to the microwave detection cavity. This also prevents the modes from the two cavities from hybridizing, which would make optimizing and tuning the individual modes challenging. This coupling scheme will result in losses from coupling the transmission line to the detection cavity, as well as ohmic losses from the transmission line. We define the total efficiency from this step as 
\begin{equation}
    \eta_0=\tilde\eta_0\kappa^{dc}
\end{equation}

\noindent where $\tilde\eta_0$ is the efficiency of the transmission line and $\kappa^{dc}$ is the coupling efficiency of the detection cavity. With low-loss superconducting cables, we expect that the efficiency of the transmission line will be dominated by the loss in the circulator. Estimating 1 dB loss in the circulator, we expect $\tilde \eta_0\sim0.8$. The coupling efficiency is given by
\begin{equation}
    \kappa^{dc}= \frac{1}{1+\beta}
\end{equation}

\noindent and has a maximum value of $1/2$, achieved for the critical coupling condition $\beta = 1$.

Once the axion-converted photon has reached the detection cavity, it must couple to the atomic beam and induce a Rydberg transition. In Sec.~\ref{sec: coupling} and Sec.~\ref{sec: beamsource}, we described the requirements for $Q_\text{det}$ and $v_L$ that enable minimal efficiency loss in the interactions between the Rydberg atoms and the detection cavity photons. When these conditions are met, the efficiency for this first step is $\eta_1 \sim 1$. 

After the Rydberg atoms have exited the detection cavity after the appropriate interaction time, their final states are probed using selective field ionization. SFI is a standard technique used for identifying Rydberg states and relies on the fact that the field needed to ionize a Rydberg state depends highly on the state's quantum number~\cite{gallagher1994}. In practice, this is achieved by using a time-varying electric field pulse that ramps up to the ionization field, such that Rydberg states with different ionization energies are ionized at different times~\cite{tada2002}. These pulsed sequences are commonly used to resolve multiple different states in an atomic ensemble at different times. However, even if we do not need to resolve different states, using a pulsed electric field provides more control over how our starting states will evolve in the Stark basis, preventing mixing between Rydberg states with similar energy levels. After ionization occurs, the ionized electrons are detected by a charged particle detector such as a channel electron multiplier (CEM) or microchannel plate (MCP). 

Efficiency losses in the readout stage can primarily arise from two sources: SFI deadtime ($\eta_2$) and electron detection ($\eta_3$). SFI deadtime is caused by atoms entering or exiting the ionization region after the pulse has already started. These atoms will not experience the intended pulse sequence and may not ionize properly. Short pulse sequences are less vulnerable to this deadtime effect. Optimizing the shape of the electric field ramp is an effective way to decrease the time needed for SFI: shaped pulse techniques have been used to reliably distinguish $S$ and $P$ Rydberg states in alkali atoms with pulse lengths typically $<$1 $\upmu$s~\cite{tada2002, gurtler2004, Gregoric2020}. This level of deadtime is negligible compared to the overall repetition rate of the single-photon detector ($\sim$10 $\upmu$s), so the ionization efficiency is $\eta_2 \sim 1$. Efficiency after ionization is set by the efficiency of the device used to collect the ionized electrons. For instance, electron collection with a channel electron multiplier can reasonably achieve $\eta_3>$ 0.8 ~\cite{Maioli2005, Shibata2003}. 

 A summary of efficiencies at various steps and associated technical assumptions can be found in Table~\ref{table: efficiency}. The resulting total efficiency is $\eta=\prod_i\eta_i \sim 0.4$. 

\begin{table}[h]\centering
\caption{Best-case efficiencies for various loss mechanisms in signal transmission and single-photon detection. }\label{table: efficiency}
\begin{tabular}{>{\centering\arraybackslash}p{3cm} c >{\centering\arraybackslash}p{3cm}} \hline\hline
Loss Mechanism &  Efficiency  & Technical Assumptions \\
  \hline
transmission line dissipation & $\tilde\eta_0\sim0.8 $ &  low-loss cables and circulator\\
\hline
coupling to transmission line  &  $\kappa^{dc}=1/2$ &  critical coupling\\
\hline
atom-microwave coupling & $\eta_1\sim 1$ & strong coupling \\
\hline
selective ionization & $\eta_2\sim 1$ & SFI pulse length $< 1\ \upmu$s \\
\hline
electron detection & $\eta_3>0.8$ & channel electron multiplier\\
\hline\hline
\end{tabular}
\end{table} 

\subsection{Total noise rate \label{sec: noise}}
Noise in single-photon detection in the 10-100 GHz range is expected to be dominated by BBR~\cite{Lamoreaux2013}. 
At temperature $T$ and frequency $\nu_a$, the photon occupation number $(\bar{n}_T$) of thermal photons is given by
\begin{align}
  \bar{n}_T &= \frac1{e^{h\nu_a/k_BT}-1} \label{eq: nthermal}.
\end{align}

\begin{figure} 
\centering
    \includegraphics[width=\linewidth]{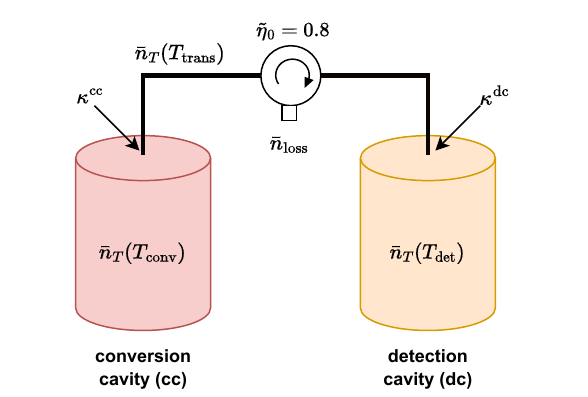}
    \caption{Signal and noise flow between the conversion cavity and detection cavity. The conversion cavity is at temperature $T_\text{conv}$ and the detection cavity is at temperature $T_\text{det}$. Photons from the conversion cavity flow to the detection cavity via a low-loss transmission line at temperature $T_\text{trans}$. A circulator with one port terminated by a 50 $\Omega$ resistor (shown as a block) is placed between the two cavities to prevent photon flow in the reverse direction.
    }
    \label{fig: noisemodel}
\end{figure}

BBR noise in the detector arises from thermal photons occupying the cavities and transmission line. In Fig.~\ref{fig: noisemodel}, we show a schematic of the two-cavity system connected by a transmission line. The transmission line is made non-reciprocal by inserting a circulator with one port terminated by a 50 $\Omega$ resistor.

In order to determine the system noise, we track the flow of thermal photons in the system. We start by considering coupling of photons into the transmission line, which is parameterized by $\kappa^\text{cc}$. The photon occupation in the transmission line at the output of the conversion cavity is given by
\begin{equation}
    \bar n_{T,\text{output}}^{cc}=\bar n_T(T_\text{conv})\kappa^{cc} + \bar n_{T}(T_\text{trans})(1-\kappa^{cc}),
\end{equation}
\noindent where the first term represents thermal photons coupled out of the cavity and the second term accounts for thermal photons in the transmission line that reflect off the cavity.

The circulator introduces efficiency losses that replace a fraction of the input with thermal noise from the circulator, assumed to be at the same temperature as the transmission line such that $\bar{n}_\text{loss} = \bar{n}_{T}(T_\text{trans})$. The coupling between the circulator loss and the input signal can be modeled as a beam-splitter~\cite{Pozar, Wang2021}, in which case the number of thermal photons incident on the detection cavity is given by
\begin{equation}
     \bar n_{T,\text{input}}^{dc}= \bar n_{T,\text{output}}^{cc}\tilde \eta_0 + \bar n_\text{loss}(1-\tilde \eta_0). \label{eq: noise_dcin}
\end{equation}

Finally, these noise photons couple into the detection cavity, which also contains its own population of thermal photons. The input noise photons are coupled into the detection cavity with efficiency $\kappa^{dc}$, while the thermal photons originating in the detection cavity are coupled out of the cavity with efficiency $1-\kappa^{dc}$, leading to a total thermal photon population of
\begin{equation}\label{eq:ndc}
\bar n^{dc}_{T}= \bar n_{T,\text{input}}^{dc}\kappa^{dc} + \bar n_T(T_\text{det})(1-\kappa^{dc}). \end{equation}

We have performed this derivation in the general case where the detection cavity and conversion cavity can have different temperatures. When estimating the noise rate of the experiment, we will assume that the two cavities, the transmission line, and the circulator are approximately thermalized to the same temperature $T$. In this case, the occupation number of the detection cavity is simply
\begin{equation}\label{eq:ndc2}
\bar n^{dc}_{T} \approx \bar n_T(T).
\end{equation}

So far, we have considered the occupation number of thermal photons with frequency precisely equal to the detection cavity's resonant frequency. However, thermal photons with frequencies across the detection cavity's bandwidth can all couple into the same cavity mode, increasing the number of noise counts ultimately coupled out of the cavity by the single-photon detector. Then, the rate of thermal photons read out of the detection cavity, $R_T$, is
\begin{equation}\label{eq: rt}
    R_T \approx \frac{2\pi \nu_a}{Q_\text{det}}\bar n^{dc}_T = 2\pi\Delta \nu_\text{det}\bar n^{dc}_T .
\end{equation}

Although thermal photons are the primary source of noise for the single-photon detector, noise can also enter the single-photon detector during the final state readout. This readout noise would be dominated by dark counts in the electron detector used for SFI. If, for example, a channel electron multiplier (CEM) from Sjuts is used for electron detection, the dark count rate is expected to be $<$0.02 ct/s~\cite{sjuts}. 

Some of these electron dark counts can be discarded using timing information. With pulsed SFI, the time that the electric field reaches the Rydberg ionization threshold is known. If a count shows up when ionization is impossible, it can be identified as a dark count. Conservatively, we estimate that these timing cuts can remove 50\% of electron dark counts without affecting the detection efficiency, corresponding to $R_\text{RO} \approx 0.01$ ct/s. With careful characterization of the SFI pulses and corresponding electron arrival times, much better dark count rejection may be possible.   

\begin{figure}\centering
    \includegraphics[width=\linewidth]{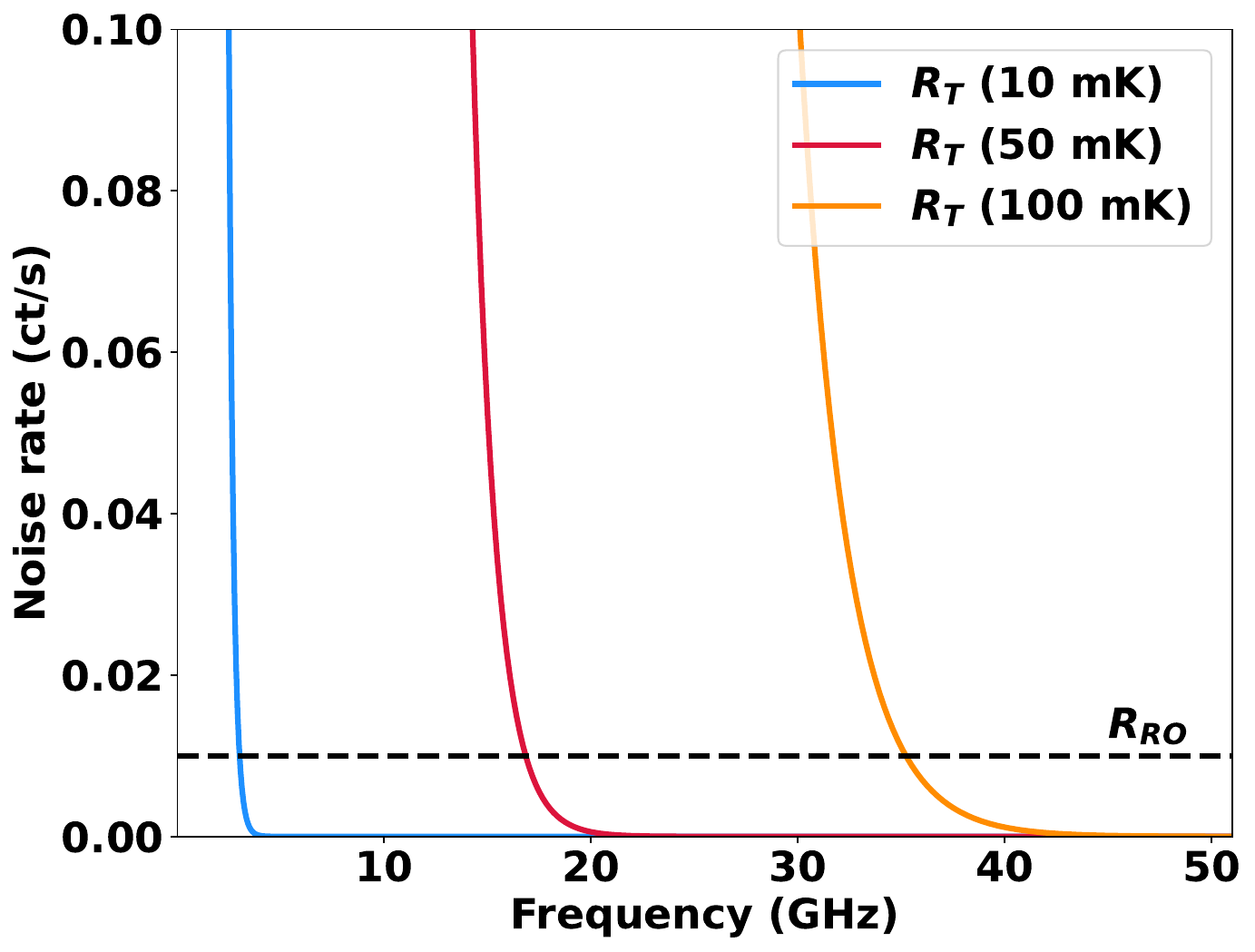}
    \caption{Rate of thermal photons read out of detection cavity $R_T$ vs. frequency, plotted for operating temperatures 10 mK, 50 mK, and 100 mK. The dashed line indicates a conservative estimate for the readout noise from electron detection $R_\text{RO}$.}
    \label{fig: noise}
\end{figure}

The noise rates for different noise sources in the experiment are estimated in Fig.~\ref{fig: noise}. $R_T$ is calculated from Eq.~\ref{eq: rt} for several operating temperatures $T$, assuming $Q_\text{det} = 10^6$ and the efficiencies from Table~\ref{table: efficiency}. $R_\text{RO}$ is estimated as 0.01 ct/s. The total noise rate in the single-photon detector is given by
\begin{equation}
    R_\text{sys} = R_T + R_\text{RO}.\label{eq: noisesum}
\end{equation}

Single-photon detectors perform best above the frequency where the noise contribution from thermal photons is negligible compared to the readout noise. From Fig.~\ref{fig: noise}, we can see the frequency at which thermal photons cease to be the dominant background: 35 GHz at 100 mK, 17 GHz at 50 mK, and 3 GHz at 10 mK. 

In many cases, the quantum measurement noise from linear amplifiers is much larger than 0.01 ct/s, and single-photon detectors will outperform linear amplifiers even outside of this ideal operating range. Detailed comparisons of single-photon detectors and linear amplifiers are discussed in Appendix~\ref{app: comparison} and indicate single-photon advantage above 15 GHz even when operating at 100 mK.

\section{Sensitivity}\label{sec: sensitivity}
We now estimate the axion sensitivity of a haloscope experiment that uses the single-photon detector described in the prior sections. We must make several assumptions regarding the overall haloscope design. Since single-photon detectors read out from a haloscope's conversion cavity, they can be incorporated into many different haloscope designs. Rather than proposing a specific haloscope for use with the single-photon detector, we explore the performance of a variety of haloscopes.

The sensitivity of an arbitrary haloscope is improved by a factor of up to 10$^4$ by switching from linear amplifier readout to single-photon detector readout, as long as the haloscope operates at a frequency and temperature where thermal photon counts are negligible. We describe the scan rate enhancement calculation in more detail in Appendix~\ref{app: comparison}. Here, we focus on calculating the axion sensitivity for several specific haloscope designs incorporating single-photon detectors. We begin with a current-generation haloscope as a benchmark and then incorporate a variety of proposed design improvements that could define the next generation of haloscopes. 

In Sec.~\ref{sec: concept}, we derived the scan rate of an axion search using a single-photon detector. Now that we have an estimate of our single-photon detector's efficiency and an expression for its noise rate (Eq.~\ref{eq: noisesum}), we can use Eq.~\ref{eq: scan_sp} to calculate the scan rate of a haloscope that uses this single-photon detector.

Several experimental parameters of the haloscope and single-photon detector that enter into the scan rate are kept constant. The haloscope design parameters $V_0$ and $C_{mn\ell}$ are set approximately equal to the operational parameters of HAYSTAC, a currently-running haloscope experiment that searches for axions around 5 GHz~\cite{Jewell2023}. $Q_\text{det}$ is set to $10^6$ based on the design constraints discussed in Sec.~\ref{sec: coupling} and Sec.~\ref{sec: beamsource}. The total efficiency is estimated in Sec.~\ref{sec: eff} as 0.4 and $\beta$ is set to 1 for critical coupling out of the conversion cavity. As in Sec.~\ref{sec: noise}, $R_\text{RO}$ is conservatively set to 0.01 ct/s. We summarize the design parameters kept constant across all sensitivity projections in Table \ref{table: parameters}.

\begin{table}[h]\centering
\caption{Parameters shared between all haloscope designs examined in the sensitivity study.}\label{table: parameters}
{
\begin{tabular}{c c c} \hline\hline
Parameter &  Value & Description\\
\hline
$V_0$ & 1.5 L & volume at 5 GHz\\
\hline
$C_{mn\ell}$ & 0.5 & form factor\\
\hline
$Q_\text{det}$ & $10^6$ & detection cavity Q\\
\hline
$\beta$ & 1 & coupling factor\\
\hline
$\eta$ & 0.4 & total efficiency\\
\hline
$R_\text{RO}$ & 0.01 ct/s & SFI noise rate\\
\hline\hline
\end{tabular}
}
\end{table} 

We consider four different haloscope design scenarios. These scenarios differ from each other in four ways: the conversion cavity quality factor $Q_\text{conv}$, the frequency-volume scaling relation $V(\nu_a)$, the magnetic field strength $B_0$, and the operating temperature $T$.  Scenario A represents the most conservative case, in which the single-photon detector is attached to a current-generation haloscope and the apparatus operates at 100 mK. Scenario B improves over Scenario A broadly across several dimensions: moderately increasing the magnetic field, improving $Q_\text{conv}$ by a factor of ten, and reducing the operating temperature to 50 mK. Scenario C considers the case of a small-bore, high B-field magnet operating in the upper end of our proposed frequency range, and also features modest improvements to $Q_\text{conv}$. Finally, in Scenario D we allow the conversion cavity volume to be independent of frequency, but use only conservative parameters for the magnetic field and quality factor. 

\begin{figure*}[ht]\centering
    \begin{subfigure}[b]{0.495\linewidth}\caption{}
    \includegraphics[width=0.9\textwidth]{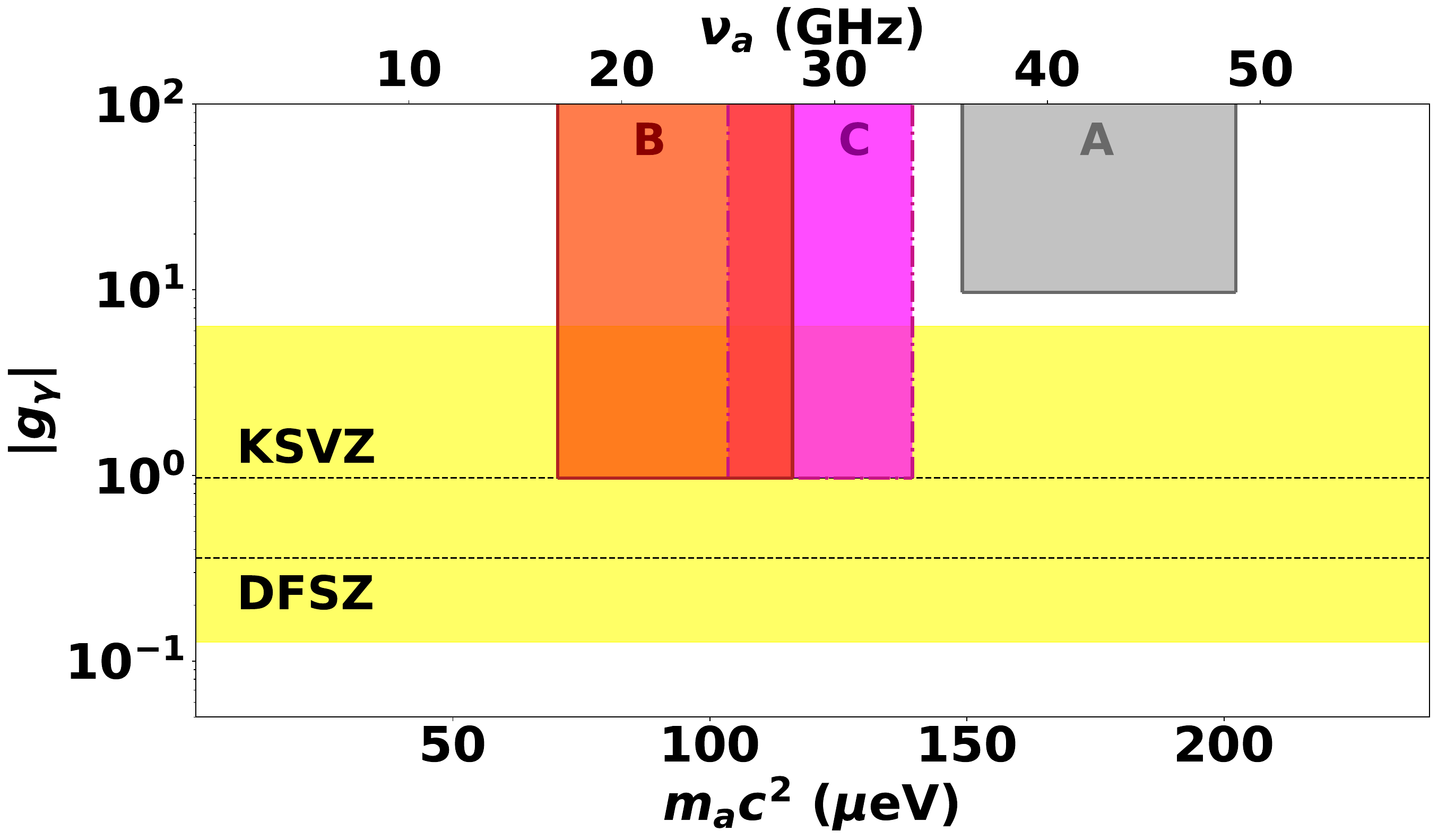}\label{fig: sens_a}
    \end{subfigure}
    \hfill
    \begin{subfigure}[b]{0.495\linewidth}\caption{}
    \includegraphics[width=0.9\textwidth]{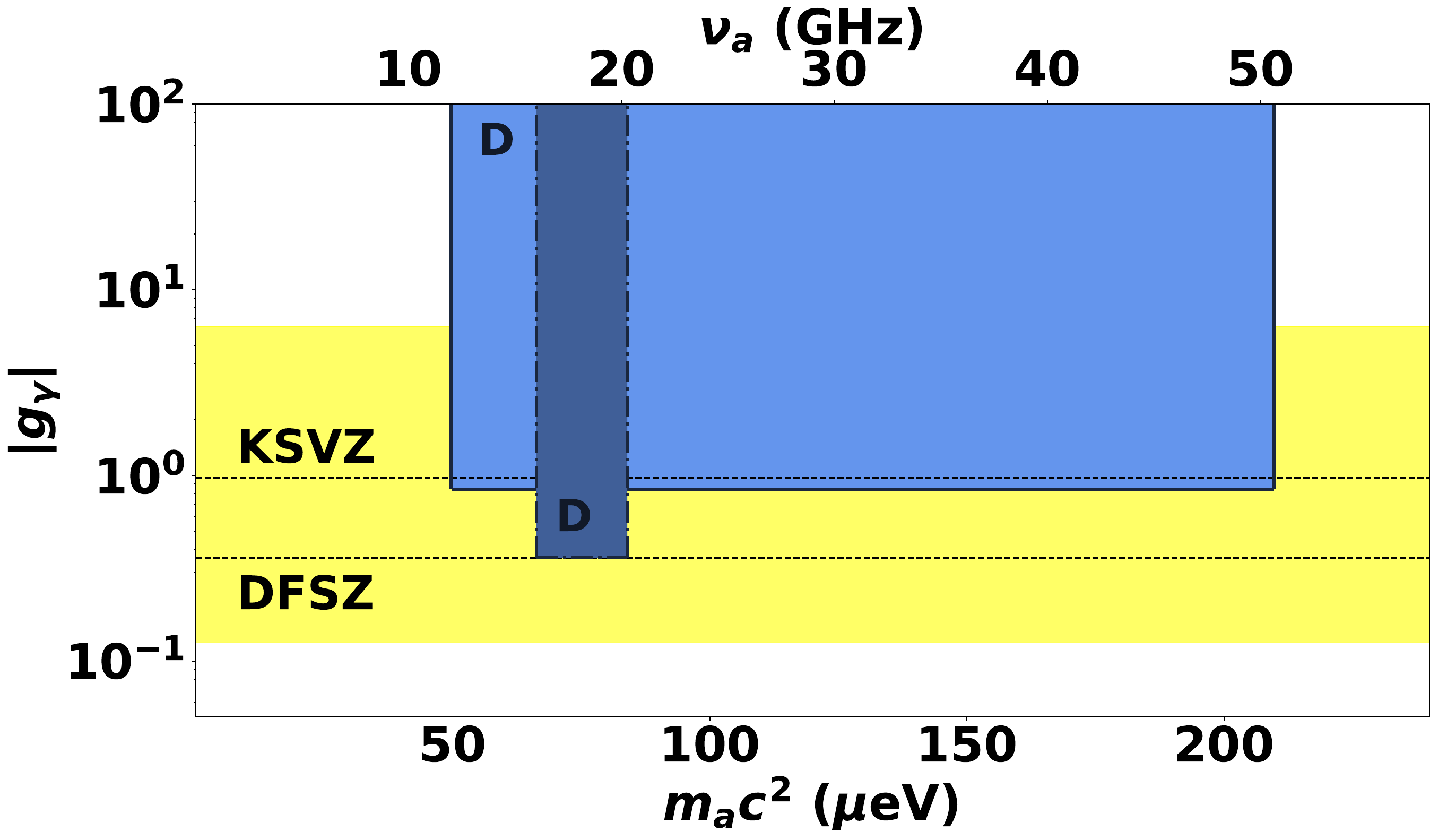}\label{fig: sens_b}
    \end{subfigure}
    \caption{Axion sensitivity for a haloscope experiment using single-photon detection. Shaded areas show the parameter space that can be covered in five years of operation. Shown in (a) are three scenarios using standard microwave haloscope cavities: A is a conservative design representing the current generation of haloscopes, B is a more ambitious design representing the next generation of haloscopes, C is a design using a small-bore magnet with a high B-field. Shown in (b) are two different searches using a non-standard cavity whose frequency is independent of the cavity size, where the narrower exclusion is taken with a longer integration time at each tuning in order to achieve DFSZ sensitivity. Experimental parameters shared between the scenarios are given in Table \ref{table: parameters} and experimental parameters differing between the scenarios are given in Table \ref{table: scenarios}. KSVZ~\cite{Kim1979,SHIFMAN1980} and DFSZ~\cite{DINE1981,Zhitnitsky1980} benchmark models are shown with dashed lines while the entire QCD band is highlighted in yellow~\cite{Diluzio2020modelband}.}
    \label{fig: exclusion}
\end{figure*}

We begin the exploration of these scenarios by discussing the quality factor of the conversion cavity. Improving $Q_\text{conv}$ increases the magnitude of the haloscope signal but poses a challenge due to the strong magnetic fields surrounding the conversion cavity that degrade the quality factor. Typical unloaded quality factors of current haloscope experiments are on the order of $10^4$, so with critical coupling ($\beta = 1$) their loaded quality factors are also on the order of $10^4$ ~\cite{admx2021, Jewell2023, CAPP2021}. Cavities coated with superconducting materials have shown higher quality factors on the order of $10^5$ even at magnetic fields up to 8 T~\cite{Posen2022, Alesini2019, Ahn2022}. High values of $Q_\text{conv}$ are particularly important when using single-photon detectors for readout, because single-photon detectors do not have frequency resolution within the conversion cavity bandwidth. While a linear-amplifier-based haloscope can take advantage of lower values of $Q_\text{conv}$ to perform a broadband measurement across several frequencies at once, we must develop new approaches when measuring with single-photon detectors.

Next, we consider the conversion cavity geometry. The current generation of haloscope experiments searches for axions at frequencies below 6 GHz. Operating an axion search in the 10-50 GHz frequency range will require adapting existing conversion cavity designs or developing new approaches altogether. The frequency dependence of the conversion cavity volume is related to the resonant frequency of the available cavity modes at tuning step. We approximate the typical frequency variation by choosing a benchmark frequency and volume and then scaling by the resonance as discussed in Sec.~\ref{sec: advantage}:
\begin{equation}
    V = V_0 \left(\frac{\nu_0}{\nu_a}\right)^3.
\end{equation}
 
 Some proposed experimental designs seek to eliminate the frequency dependence of the conversion cavity volume altogether, such as dielectric haloscopes and plasma haloscopes~\cite{Caldwell2017, Lawson2019}. The broad bandwidth of dielectric haloscopes ($\sim$10 MHz) may be challenging to integrate with narrow-band single-photon detection. However, the bandwidth of plasma haloscopes is similar to the bandwidth of a standard haloscope cavity: plasma haloscope cavities are projected to be capable of achieving unloaded quality factors on the order of $10^4$~\cite{ALPHA2023}. Since this is a promising area of axion search development, we consider volume independence in one of our scenarios.

The magnetic field used for a haloscope also has a large impact on the ultimate sensitivity, as the scan rate scales with $B_0^4$. Current haloscope experiments already use a variety of magnetic fields. For instance, HAYSTAC uses a magnetic field of 8 T~\cite{Jewell2023}. However, magnetic fields up to 18 T have been used by the CAPP axion experiment~\cite{Lee2022}. Development of magnets for haloscope experiments requires not only producing a large magnetic field, but also maintaining that magnetic field over the volume occupied by the conversion cavity. Typically, higher field magnets require smaller bores: one design using superconducting inserts has achieved 32 T fields with a 3.4 cm bore~\cite{SCM32T}. While this bore size could not hold an experiment like HAYSTAC, which has a diameter of 12.7 cm at 5 GHz, at higher frequencies the cavity diameter would decrease as $\nu_a^{-1}$, allowing such a magnet to be used at frequencies above 20 GHz.

\begin{table}[h]\centering
\caption{Parameters used to specify four different scenarios of a single-photon detector connected to a haloscope.}\label{table: scenarios}
{
\begin{tabular}{c c c c c} \hline\hline
Scenario &  $Q_\text{conv}$ & $V(\nu_a)$ & $B_0$ (T) & $T$ (mK)\\
\hline
conservative (A)& $1\mathrm{e}4$ & $\propto \nu_a^{-3}$ & 8& 100\\
high $Q_\text{conv}$(B) & $1\mathrm{e}5$ & $\propto \nu_a^{-3}$& 12& 50\\
high $B_0$ (C) & $2\mathrm{e}4$ &$\propto \nu_a^{-3}$ & 32& 50\\
volume-independent (D) &$1\mathrm{e}4$ & $\propto$ 1& 8& 50\\
\hline\hline
\end{tabular}
}
\end{table}

The axion sensitivity of these four hypothetical experiments for five years of data-taking is presented in Fig.~\ref{fig: exclusion} with the main parameters for each scenario summarized in Table~\ref{table: scenarios}. As shown in scenario A, searches for axion-like particles can be performed across over $\sim$40 $\upmu$eV of parameter space using the single-photon detector with existing haloscope technology. When the single-photon detector is combined with improvements in the magnetic field strength and conversion cavity design, as in scenarios B and C, QCD axion searches across $\sim$40 $\upmu$eV of parameter space become feasible. If the frequency-volume scaling can be broken, single-photon detectors allow for thorough QCD axion searches even with modest magnetic field and conversion cavity $Q$, enabling either a KSVZ axion search across the entire 40 $\upmu$eV--200 $\upmu$eV mass range or a DFSZ search across $\sim$20 $\upmu$eV of parameter space. Across all of these scenarios, single-photon detectors enhance the discovery potential of their companion haloscopes, opening up regions of parameter space otherwise inaccessible to those haloscopes.
 
\section{Summary and Conclusion \label{sec: conclude}}
Searching for axions with a cavity haloscope experiment requires sensitivity to an extremely low photon flux, especially for higher masses above 40 $\upmu$eV (10 GHz). One way of approaching this challenge is to replace standard linear amplifiers with single-photon detectors, which are not subject to the quantum measurement noise that dominates linear amplifiers at high frequencies and can achieve much higher signal-to-noise ratios as a result.

In this paper, we presented a scheme for a Rydberg-atom-based single-photon detector that can be used for axion searches in the mass range 40--200 $\upmu$eV. This single-photon detector can be connected to a wide variety of haloscope designs, replacing standard linear-amplifier-based readout schemes and delivering scan rate enhancement factors of 10$^4$. Reaching the full potential of single-photon detection with Rydberg atoms will require producing high-$Q$ cavities for atom-photon coupling and developing techniques for working with Rydberg atoms in cryostats with operating temperatures below 100 mK. When Rydberg-atom-based single-photon detector readout is combined with projected improvements for the next generation of magnetic fields and axion-photon conversion cavities, haloscope searches across ~40 $\upmu$eV ($\sim$10 GHz) of QCD axion parameter space become feasible.

\begin{acknowledgments}
We would like to thank Karl Berggren, Karl van Bibber, Charles Brown, Aaron Chou, Steve Lamoreaux, Konrad Lehnert, Mark Saffman, and the HAYSTAC collaboration for helpful discussions. We also thank Olivia Aspegren, Sophia Getz, Annie Giman,  Gabe Hoshino, Elizabeth Ruddy, and Arushi Srivastava for their assistance. This material is based upon work supported by the U.S. Department of Energy, Office of Science, Office of High Energy Physics through the QuantiSED program, under contract number DE-AC02-07CH11359.
\end{acknowledgments}

\appendix

\section{Comparison to Linear Amplifiers \label{app: comparison}}
We compare the performance of our single-photon detector to the performance of standard linear amplifier readout by calculating the scan rate enhancement. We define the scan rate enhancement $E$ as the ratio of scan rates for axion searches conducted with a single-photon detector and a linear amplifier attached to the same haloscope. Dividing Eq.~\ref{eq: scan_sp} by Eq.~\ref{eq: scan_la} under the approximation $\Delta \nu_\text{det} \approx \Delta \nu_a$,

\begin{equation}
    E = \frac{5}{4} \frac{\Delta \nu_a}{\Delta \nu_\text{conv}} \frac{R_\text{sig,sp}^2}{R_\text{sig,la}^2} \frac{N_\text{sys,la}^2 \Delta \nu_a}{R_\text{sig,sp} + R_\text{sys,sp}}.
\end{equation}

We simplify this expression in a few ways. First, we drop the factor of 5/4. Then, we rewrite $\Delta \nu_a/{\Delta \nu_\text{conv}}$ in its equivalent form $Q_\text{conv}/Q_a$. Next, we assume that the two signal rates are approximately equal and drop the subscript. This is equivalent to assuming that the single-photon detector and linear amplifier readout have the same efficiency, which is reasonable according to Sec.~\ref{sec: eff}. Finally, we set $N_\text{sys,la} \approx 1$ for a linear amplifier operating at the SQL, ignoring the relatively small contribution of thermal noise.

Using these simplifications and writing the single-photon noise rate as in Eq.~\ref{eq: noisesum},

\begin{equation}
    E \approx \frac{Q_\text{conv}}{Q_a} \frac{\Delta \nu_a}{R_\text{sig} + R_T + R_\text{RO}}.
\end{equation}

At lower frequencies, the noise affecting the scan rate enhancement primarily comes from BBR. The exact frequency where the single-photon detector starts to outperform the linear amplifier depends on the operating temperature and the ratio of the conversion cavity $Q$ to the axion $Q$, consistent with the estimates made in \cite{Lamoreaux2013}. For low enough temperature and high enough frequency, the BBR noise becomes negligible and the other sources of noise in the single-photon detector take precedence: shot noise in the signal counts and dark counts from the readout mechanism. 

At higher frequencies, the dominant source of noise comes from shot noise in the single-photon detector. Although the readout dark count rate can be improved through use of low-noise electron detectors, shot noise comes directly from the statistics of searching for axions by counting photons. Since linear amplifier readout does not have shot noise, this reduces the maximum scan rate enhancement: increasing $R_\text{sig}$ increases the SNR of the single-photon detector by a factor of $\sqrt{R_\text{sig}}$ but increases the SNR of the linear amplifier by a factor of $R_\text{sig}$. If all other sources of noise are negligible and $Q_\text{conv}/Q_a\approx 0.01$, single-photon detectors are favorable for $R_\text{sig} <$ 100 ct/s and can achieve a scan rate enhancement of 10$^3$ for $R_\text{sig} <$ 1 ct/s. 

Signal rates that could significantly reduce the scan rate enhancement achievable with a single-photon detector are well in excess of the signal rates achievable with haloscopes operating above 10 GHz. For instance, the HAYSTAC experiment would see 0.3 ct/s from KSVZ axions when operating around 5 GHz~\cite{HAYSTAC2021}. As indicated in Eq. \ref{eq: scaling}, the signal power is proportional to the inverse cube of the frequency, so reaching 1 ct/s at 10 GHz and 50 GHz would require improving the baseline signal power of that haloscope by a factor of 25 and 25000 respectively. 

 \begin{figure}\centering
    \includegraphics[width=\linewidth]{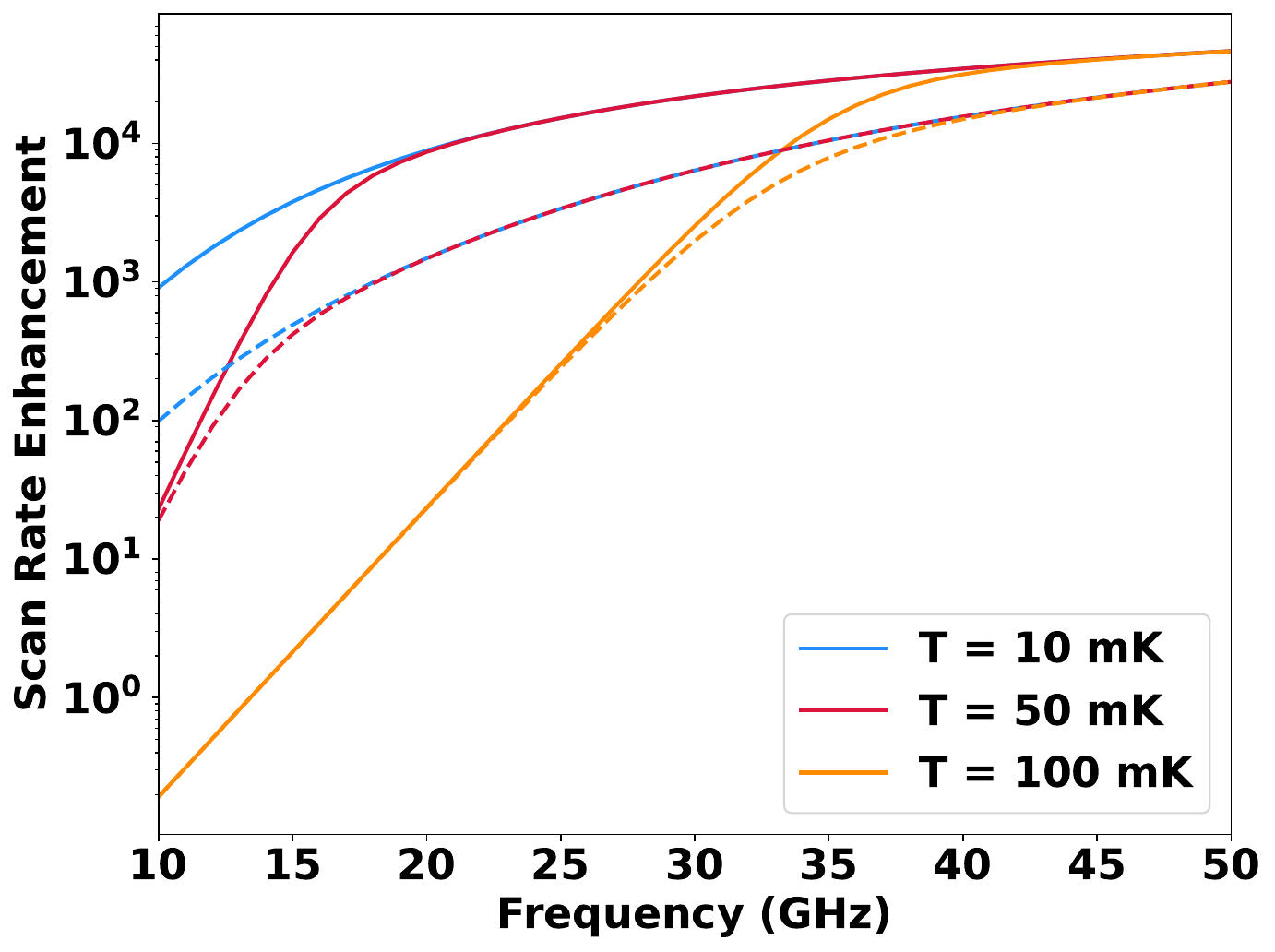}
    \caption{Scan rate enhancement for a haloscope using the proposed single-photon detector for readout instead of a SQL-limited linear amplifier, shown for a readout dark count rate of 0.01 ct/s and operating temperatures of $T=$ 10 mK, 50 mK, and 100 mK. Solid lines show the enhancement for $R_\text{sig}(\text{10 GHz}) = 0.1$ ct/s, while dashed lines show the enhancement for $R_\text{sig}(\text{10 GHz}) = 1$ ct/s.}
    \label{fig: enhancement}
\end{figure}

The scan rate enhancement as a function of frequency is shown in Fig.~\ref{fig: enhancement} for several different values of $T$ and $R_\text{sig}(\text{10 GHz})$. We estimate $Q_\text{conv}/Q_a\approx 0.01$ and $R_\text{RO} \approx 0.01$ ct/s. At 10 mK, BBR noise is negligible across the entire frequency range. At 50 mK, BBR noise can be seen until around 20 GHz, but the single-photon detector still outperforms the linear amplifier across the entire frequency range. At 100 mK, BBR noise is present until 40 GHz, and the single-photon detector becomes favorable to the linear amplifier around 15 GHz. In the absence of BBR, the scan rate enhancement becomes near-constant with frequency, saturating around $5\times10^4$ for signal rates of 0.1 ct/s and around $10^4$ for signal rates of 1 ct/s. 

\bibliography{revisedbib}
\end{document}